\documentclass[reprint,showkeys,amsmath,prd,floatfix,superscriptaddress,nofootinbib]{revtex4-1}
\usepackage{adjustbox}
\usepackage{graphicx}
\usepackage{hyperref}
\usepackage{xspace}
\usepackage{xcolor}
\usepackage{siunitx}
\graphicspath{{pics/}}
\newcommand{\xFitter}{\texttt{xFitter}\xspace}%
\def\al{\alpha}
\def\de{\delta}
\def\ga{\gamma}
\def\dif{\textrm{d}}
\def\defeq{=}

\def\alS{\alpha_S}
\def\GeV{\ \text{GeV}}
\def\Beta{\mathcal{B}} 
\newcommand{\figref}[1]{Fig.~\ref{fig:#1}}
\newcommand{\Figref}[1]{Figure~\ref{fig:#1}}
\newcommand{\tblref}[1]{Table~\ref{tbl:#1}}
\newcommand{\eqnref}[1]{Eq.\eqref{eq:#1}}
\let\citeA\cite
\renewcommand{\cite}[1]{~\citeA{#1}}
\begin{document}
\title{Parton Distribution Functions of the Charged Pion Within The \xFitter Framework}
\author{Ivan Novikov}
\email{ivan.novikov@desy.de}
\affiliation{Joint Institute for Nuclear Research, Joliot-Curie 6, Dubna, Moscow region, Russia, 141980}
\affiliation{Deutsches Elektronen-Synchrotron (DESY), Notkestrasse 85, D-22607 Hamburg, Germany}
\author{Hamed Abdolmaleki}
\affiliation{School of Particles and Accelerators, Institute for Research in Fundamental Sciences (IPM), P. O. Box 19395-5531, Tehran, Iran.}
\author{Daniel Britzger}
\affiliation{Max-Planck-Institut f\"ur Physik, F\"ohringer Ring 6, D-80805 M\"unchen, Germany}
\author{Amanda Cooper-Sarkar}
\affiliation{Particle Physics, Denys Wilkinson Bdg, Keble Road, University of Oxford, OX1 3RH Oxford, UK}
\author{Francesco Giuli}
\affiliation{University of Rome Tor Vergata and INFN, Sezione di Roma 2, Via della Ricerca Scientifica 1,00133 Roma, Italy}
\author{Alexander Glazov}
\email{alexander.glazov@desy.de}
\affiliation{Deutsches Elektronen-Synchrotron (DESY), Notkestrasse 85, D-22607 Hamburg, Germany}
\author{Aleksander Kusina}
\affiliation{Institute of Nuclear Physics Polish Academy of Sciences, PL-31342 Krakow, Poland}
\author{Agnieszka Luszczak}
\affiliation{T. Kosciuszko Cracow University of Technology, PL-30-084, Cracow, Poland}
\author{Fred Olness}
\affiliation{Southern Methodist University, Department of Physics, Box 0175 Dallas, TX 75275-0175, United States of America}
\author{Pavel Starovoitov}
\affiliation{Kirchhoff-Institut f\"ur Physik, Heidelberg University, Im Neuenheimer Feld 227, 69120 Heidelberg, Germany}
\author{Mark Sutton}
\affiliation{Department of Physics and Astronomy, The University of Sussex, Brighton, BN1 9RH, United Kingdom}
\author{Oleksandr Zenaiev}
\affiliation{Hamburg University, II. Institute for Theoretical Physics, Luruper Chaussee 149, D-22761 Hamburg,Germany}
\collaboration{xFitter Developers' team}
\date{\today}
\begin{abstract}
%
%
%
We present the first open-source analysis of
parton distribution functions (PDFs) of charged pions 
using \xFitter, 
  an open-source QCD fit framework to facilitate PDF extraction and analyses.
Our calculations are implemented at NLO using APPLgrids generated by MCFM. 
Using  currently available Drell-Yan and photon production data, 
we find   the valence distribution is well-constrained; however, 
the considered data are not sensitive enough to unambiguously determine sea and gluon distributions.
Fractions of momentum carried by the valence, sea and gluon components are discussed, and
we compare with the JAM and GRVPI1 results.
%
\end{abstract}
\maketitle
\section*{Introduction}
The pion plays an important role in our understanding of strong interactions.
At the same time, it is a mediator of nucleon-nucleon interactions, a pseudo-Goldstone boson of dynamical chiral symmetry breaking and the simplest $q\bar q$ state in the quark-parton model of hadrons.
However, from the experimental point of view, the pion structure is currently poorly understood, especially compared to the proton.
Parton distribution functions (PDFs) are a primary theoretical construct used to describe hadron structure as it is probed in hard processes.
Much progress has been made in mapping out the parton distribution functions of the proton in the last decades\cite{proton}.
\par
On the other hand, theoretically, the pion is a simpler system than the proton.
Consequently, the pion structure has been investigated in several nonperturbative theoretical models.
Nambu-Jona-Lasinio model\cite{NJL_5,NJL_6,NJLpion},
Dyson-Schwinger equations\cite{DSE_2011,DSE_2014,DSE_2016,DSE_2018,DSE_2019,Ding:2019qlr,Ding:2019lwe} (DSE),
meson cloud model\cite{MesonCloudNonperturbativeSea},
and nonlocal chiral-quark model\cite{ChQM_2012,ChQM_2016,ChQM_2018}
make predictions about certain aspects of PDFs of the charged pion, or even allow calculating PDFs themselves.
In the lattice QCD approach first moments of the valence pion PDF have been calculated\cite{Lattice1997,Lattice2003,Lattice2016},
and direct computation of PDF has recently been achieved\cite{LatticePDF1,LatticePDF2,LatticePDF3,LatticePDF4}.
\par
Experimentally, the pion PDF is known mostly from QCD analyses of Drell-Yan (DY) and prompt photon production data\cite{Owens,Aurenche,SMRS,Wijesooriya}.
Within a dynamical approach, only the relatively well-known valence distribution is determined from DY data,
with the sea and gluon content at a very low initial scale fixed by simplifying assumptions\cite{GRVPI} or constraints of the constituent quark model\cite{GRS1997,GRS1999}.
While all modern pion PDF extractions are performed at next-to-leading order (NLO),
additional threshold-resummation corrections and their impact on the valence distribution at high $x$ have been studied\cite{SoftGluonResummation}.
In addition to DY data, a recent work by the JAM collaboration\cite{JAM} included leading neutron (LN) electroproduction data obtained from the HERA collider (as suggested in \cite{LeadingNeutronHERA}).
The latest pion PDF fit by Bourelly and Soffer\cite{FitBourelly} uses a novel parameterisation at the initial scale $Q_0$.
\par

In this analysis we approach the pion PDFs from a phenomenological
context and introduce a number of unique features which provide a
complementary perspective relative to other determinations.
In particular, the combination of DY (E615 and NA10) and prompt photon
(WA70) data provide constraints on both the quarks and gluons in our
kinematic range.
We also explore the theoretical uncertainties including variations of
the strong coupling, as well as the factorization and renormalization
scales; consequently, our PDF error bands reflect both the
experimental and theoretical uncertainties.
Our analysis uses MCFM generated APPLgrids which allow for efficient
numerical computations; additionally, we implemented modifications to
APPLgrid which allow both meson and hadron PDFs in the initial state.
This work is implemented in the publicly available \xFitter PDF
fitting framework\cite{xFitter}; as such, it is the first open-source
analysis of pion PDFs, and this will facilitate future studies of
meson PDFs as new data become available.

\par
The paper is organised as follows:
In Section \ref{sec:data} we briefly discuss the considered data.
The adopted PDF parameterisation and decomposition are described in Section \ref{sec:pars}.
Calculation of theoretical predictions is discussed in Section \ref{sec:theory}.
Section \ref{sec:errors} is devoted to the statistical treatment used in this work
and estimation of the uncertainty of the obtained PDFs.
Finally, the results of the analysis are presented and compared to results of other studies in Section \ref{sec:results}.

\section{Experimental data}\label{sec:data}
This analysis is based on Drell-Yan data from \texttt{NA10}\cite{NA10} and \texttt{E615}\cite{E615} experiments,
and on photon production data from the \texttt{WA70}\cite{WA70} experiment.
The \texttt{NA10} and \texttt{E615} experiments studied scattering of a $\pi^-$ beam off a tungsten target, with $E_\pi=194$ and $286\GeV$ in the \texttt{NA10} experiment and $E_\pi=252\GeV$ in the \texttt{E615} experiment.
The \texttt{WA70} experiment used $\pi^\pm$ beams and a proton target.
For the Drell-Yan data, the $\Upsilon$-resonance range, which corresponds to bins with $\sqrt{\tau}\in[0.415,0.484]$, were excluded from the analysis.
Here $\sqrt{\tau}\defeq m_{\mu\mu}/\sqrt{s}$, $m_{\mu\mu}$ is the invariant mass of the muon pair,
and $\sqrt{s}$ is the center-of-mass energy of pion-nucleon system.
\par
\begin{figure}[tb]
	\centerline{\includegraphics[width=8cm]{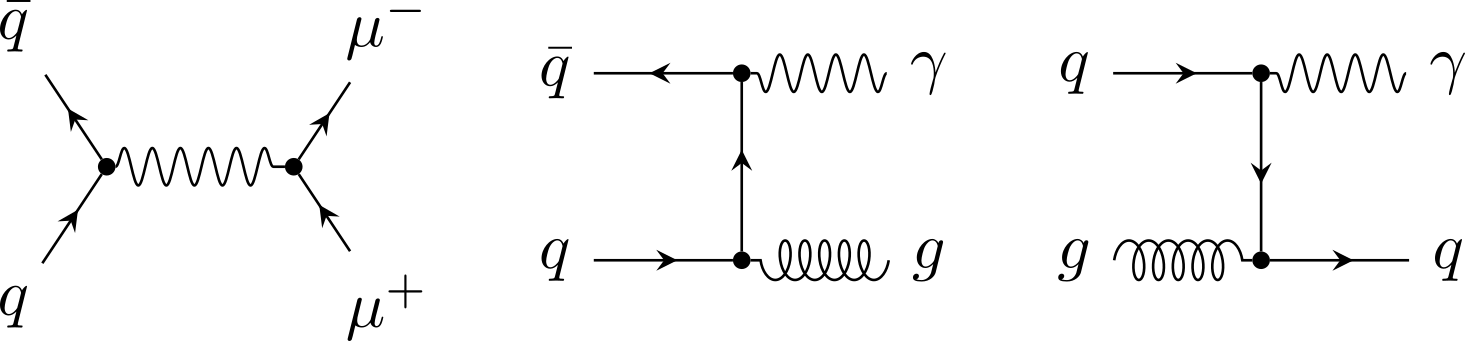}}
	\caption{\label{fig:LOdiagrams}
Leading order Feynman diagrams for the considered processes:
Drell-Yan dimuon production (left) and direct photon production (center and right).
	}
\end{figure}
Leading order Feynman diagrams for the considered processes are shown in \figref{LOdiagrams}.
The Drell-Yan data constrain the valence distribution relatively well, but are not sensitive to sea and gluon distributions.
The prompt photon production data complement the DY data by providing some sensitivity to the gluon distribution, but have smaller statistics and large uncertainties in comparison to the DY data.
Additionally, the predictions for prompt photon production have significant theoretical uncertainty, as discussed in Section \ref{sec:theory}.
\section{PDF Parameterisation}\label{sec:pars}
The $\pi^-$ PDF $xf(x,Q^2)$ is parameterized at an initial scale $Q_0^2=1.9\GeV^2$, just below the charm mass threshold $m_c^2=2.04\GeV^2$.
Neglecting electroweak corrections and quark masses, charge symmetry is assumed: $d=\bar u$, and SU(3)-symmetric sea: $u=\bar d=s=\bar s$.
Under these assumptions, pion PDFs are reduced to three distributions: total valence~$v$, total sea~$S$, and gluon~$g$:
\begin{align*}
	v&\defeq d_v-u_v=(d-\bar d)-(u-\bar u)=2(d-u)=2d_v,\\
	S&\defeq2u+2\bar d+s+\bar s=6u,\\
	g&\defeq g,
\end{align*}
which we parameterise using a generic form:
\begin{align}
	xv(x)&=A_vx^{B_v}(1-x)^{C_v}(1+D_vx^\al), \nonumber \\
        \label{eq:parm}
	xS(x)&=A_Sx^{B_S}(1-x)^{C_S}/\Beta(B_S+1,C_S+1),  \\
	xg(x)&=A_g(C_g+1)(1-x)^{C_g}, \nonumber
\end{align}
where $\Beta$ is the Euler beta function,
which ensures that the $A_S$ parameter represents the total momentum fraction carried by the sea quarks.
The $B$\nobreakdash-parameters determine the low-$x$ behavior, and $C$\nobreakdash-parameters determine the high-$x$ behavior.
Quark-counting and momentum sum rules have the following form for $\pi^-$:
\begin{align}
	\int_0^1\hskip -2mm v(x)\dif x&=2,&
	\int_0^1\hskip -2mm x(v(x)+S(x)+g(x))\dif x&=1.\label{eq:sumrules}
\end{align}
The sum rules determine the values of parameters $A_v$ and $A_g$, respectively.
The constant factors in the definitions of $v$, $S$, $g$ were chosen in such a way, that
$\langle xv\rangle, \langle xS\rangle, \langle xg\rangle$
are momentum fractions of pion carried by the valence quarks, sea quarks, and gluons, respectively
(here $\langle xf\rangle\defeq\int_0^1 xf(x)\dif x$).
\par
\begin{figure}[tb]
	\centering
	\includegraphics[width=8.6cm]{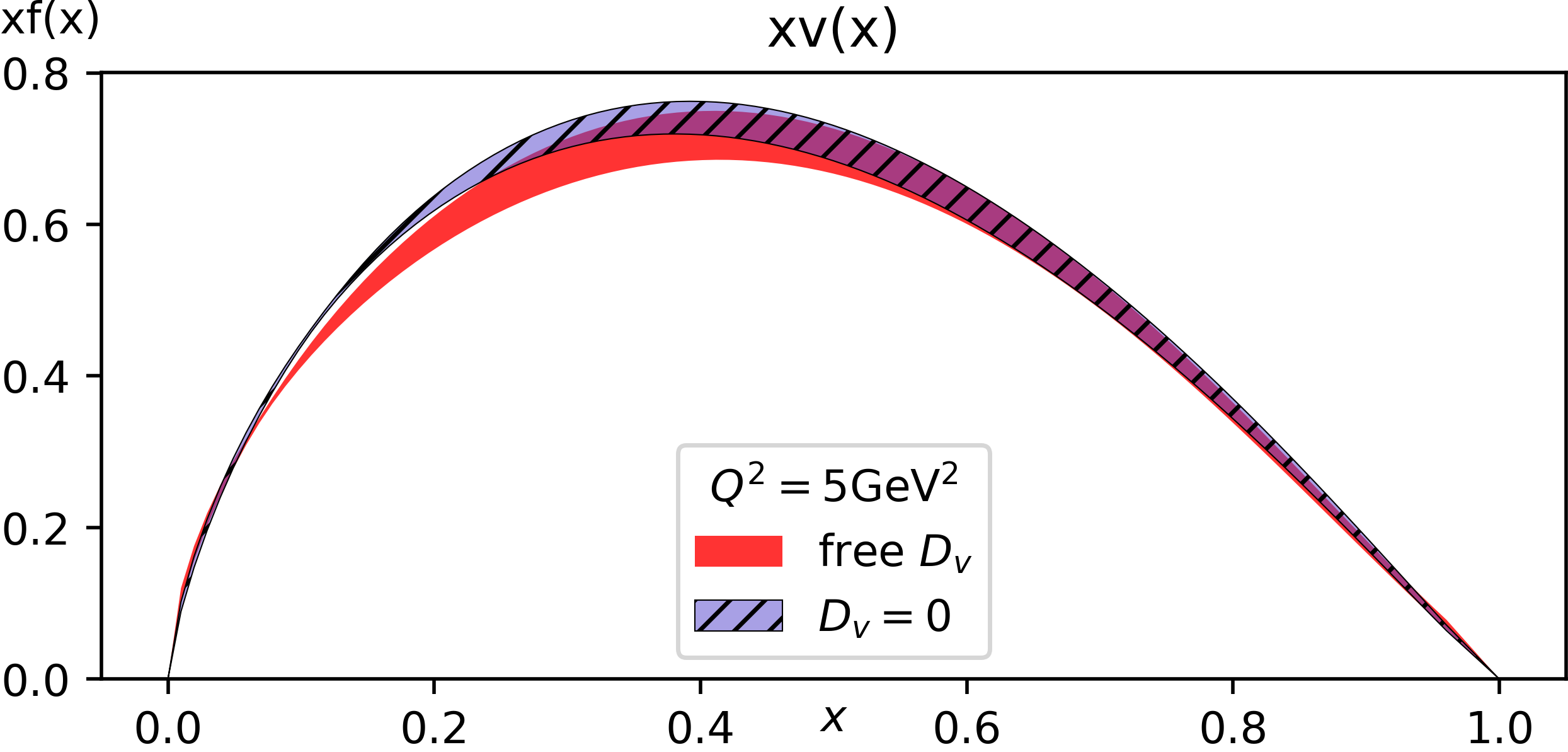}
	\caption{\label{fig:HighX}
The valence distribution when using minimal parameterisation ($D_v=0$) and the extended parameterisation with free $D_v$.
The shown uncertainty bands do not include scale variations.
	}
\end{figure}

The extension $D_vx^\al$ was introduced in $xv(x)$ to mitigate possible bias due to inflexibility of the chosen parameterisation.
This extension was omitted in the initial fits ($D_v=0$).
Afterwards, a parameterisation scan was performed by repeating the fit with free $D_v$ and different fixed values of parameter $\al$.
The scan showed that only $\al=\frac52$ has noticeably improved the quality of the fit (see \tblref{parameters} and Section \ref{sec:results} for discussion).
The additional free parameter $D_v$ changes the shape of the valence distribution only slightly (\figref{HighX}).
Similar attempts to add more parameters of the form $(1+D_vx^\al+E_vx^\beta)$ did not result in significant improvement of $\chi^2$.
The final presented results use a free $D_v$ and $\al=\frac52$.
\begin{table}[t]
	\centering
	\caption{\label{tbl:parameters}
Fitted parameter values and $\chi^2$.
The first column corresponds to the fit with $D_v=0$.
The second column shows results of the fit with free $D_v$ and $\al=\frac52$.
The uncertainties of parameter values do not include scale variations.
The valence and gluon normalization parameters $A_v$ and $A_g$ were not fitted, but were determined based on sum rules (\eqnref{sumrules}) and values of the fitted parameters.
	}
	{\hfuzz=1cm 
	\begin{tabular}{cSS}
		\hline\hline
		&{$D_v$=0}&{free $D_v$}\\
		\hline
		$\chi^2/N_{\text{DoF}}$&{\quad444/373=1.19\hspace{.5em}}&{\hspace{.5em}437/372=1.18\quad}\\
		\hline
		$A_v$                  &2.60   & 1.72\\
		$\langle xv\rangle$    &0.56   & 0.54\\
		$B_v$                  &0.75(3)& 0.63(6)\\
		$C_v$                  &0.95(3)& 0.26(13)\\
		$D_v$                  &0      &-0.93(6)\\
		$A_S=\langle xS\rangle$&0.21(8)& 0.25(9)\\
		$B_S$                  &0.5(8) & 0.3(7)\\
		$C_S$                  &8(3)   & 6(3)\\
		$A_g=\langle xg\rangle$&0.23   & 0.20\\
		$C_g$                  &3(1)   & 3(1)\\
		\hline\hline
	\end{tabular}
	}
\end{table}
\section{Cross-section calculation}\label{sec:theory}
PDFs are evolved up from the starting scale $Q_0^2$ by solving the DGLAP equations numerically using \texttt{QCDNUM}\cite{QCDNUM}.
The evolution is performed using the variable flavor-number scheme with quark mass thresholds at
$m_c=1.43\GeV$,
$m_b=4.5\GeV$.
Predictions for the cross-sections were calculated as a convolution of the evolved pion PDFs with precomputed grids of NLO coefficients and with PDFs of a proton or tungsten target.
The \texttt{APPLgrid}\cite{APPLgrid} package was used for these calculations.
The grids were generated using the \texttt{MCFM}\cite{MCFM} generator.
For Drell-Yan, the invariant mass of the lepton pair was used for the renormalization and factorisation scales, namely $\mu_R=\mu_F=m_{ll}$.
For prompt photon production, the scale was chosen as the transverse momentum of the prompt photon, namely $\mu_R=\mu_F=p_{T}(\ga)$.
\par
It was verified that the grid binning was sufficiently fine
by comparing the convolution of the grid with the PDFs used for the grid generation and a reference cross-section produced by \texttt{MCFM}.
The deviation from the reference cross-section, as well as estimated statistical uncertainty of the predictions, are an order of magnitude smaller than the uncertainty of the data.
This check was performed for each data bin.
\par
Both the evolution and cross-section calculations are performed at next-to-leading order (NLO).
For the tungsten target, nuclear PDFs from \texttt{nCTEQ15}\cite{nCTEQ15} determination were used.
In the case of a proton target, the PDFs from ref.\cite{proton-baseline} were employed.
These were also used as the baseline in the \texttt{nCTEQ15} study.
The use of another popular nuclear PDF set \texttt{EPPS16}\cite{EPPS16} was omitted because their fit had used the same pion-tungsten DY data as the present analysis.
Considering $\pi^-N$ data, \texttt{EPPS16} fitted PDFs of tungsten using fixed pion PDFs from an old analysis by GRV\cite{GRVPI}.
Nevertheless, as the \texttt{nCTEQ15}  and  \texttt{EPPS16} PDFs are comparable, within uncertainties,
this choice should not be consequential. 

\par
In the case of prompt photon production, the contribution of fragmentation photons cannot be accounted for using the described techniques.
The model used in the fit included only the direct photons.
We estimate the impact of the missing fragmentation contribution
by comparing the total integrated cross-sections computed using \texttt{MCFM} for proton-proton collision at the \texttt{WA70} energy with and without fragmentation.
The relative difference of $32\%$ is treated as the theoretical uncertainty in overall normalization of the \texttt{WA70} data.
In the run without fragmentation, Frixione isolation is used.
In the other run the fragmentation function set \texttt{GdRG\_\_LO} and cone isolation are used.
The isolation cone size parameter is $R_0=0.4$ for both cases.
\section{Statistical treatment and estimation of uncertainties\label{sec:errors}}
The PDF parameters are found by minimizing the $\chi^2$ function defined as
\begin{equation}
	\label{eq:chi2}
	\chi^2\defeq\sum_i\frac{(d_i-\tilde t_i)^2}{\left(\de^\text{syst}_i\right)^2+\left(\sqrt{\frac{\tilde t_i}{d_i}}\de^\text{stat}_i\right)^2}+\sum_\al b_\al^2,
\end{equation}
where $i$ is the index of the datapoint
and $\al$ is the index of the source of correlated error.
The measured cross-section is denoted by $d_i$,
with $\de^\text{syst}_i$ and $\de^\text{stat}_i$ being respectively the corresponding systematic and statistical uncertainties.
The $t_i$'s represent the calculated theory predictions,
and $\tilde t_i=t_i\left(1-\sum_\al\ga_{i\al}b_\al\right)$ are theory predictions corrected for the correlated shifts.
$\ga_{i\al}$ is the relative coefficient of the influence of the correlated error source $\al$ on the datapoint $i$,
and $b_\al$ is the nuisance parameter for the correlated error source $\al$.
\par
The error rescaling $\tilde\de^\text{stat}=\sqrt{\frac{\tilde t_i}{d_i}}\de^\text{stat}$ is used to correct for Poisson fluctuations of the data.
Since statistical uncertainties are typically estimated as a square root of the number of events, a random statistical fluctuation down in the number of observed events leads to a smaller estimated uncertainty, which gives such points a disproportionately large weight in the fit.
The error rescaling corrects for this effect.
This correction was only used for the Drell-Yan data.
\par
The nuisance parameters $b_\al$ are used to account for correlated uncertainties.
In this analysis the correlated uncertainties consist of
the overall normalization uncertainties of the datasets,
the correlated shifts in predictions related to uncertainties from nuclear PDFs,
and the strong coupling constant $\alS(M_Z^2)=0.118\pm0.001$.
The nuisance parameters are included in the minimization along with the PDF parameters.
They determine shifts of the theory predictions and contribute to the $\chi^2$ via the penalty term $\sum_\al b_\al^2$.
For overall data normalization, the coefficients $\ga_{i\al}$ are relative uncertainties as reported by the corresponding experiments,
and, in the case of the \texttt{WA70} data, the abovementioned additional $32\%$ theoretical uncertainty,
(listed in \tblref{data}).
\begin{table}[tb]
	\caption{\label{tbl:data}
The normalization and partial $\chi^2$ for the considered datasets.
The normalization uncertainty is presented as estimated by corresponding experiments.
In order to agree with theory predictions, the measurements must be multiplied by the normalization factor.
Deviations from 1 in the normalization factor lead to a penalty in $\chi^2$, as described in Section \ref{sec:errors}.
	}
	\begin{tabular}{ccccc}
		\hline\hline
		Experiment
		&\begin{tabular}{c}Normalization\\uncertainty\end{tabular}
		&\begin{tabular}{c}Normalization\\factor     \end{tabular}
		&$\chi^2/N_{\text{points}}$
		\\\hline
		\href{https://doi.org/10.17182/hepdata.23151   }{\texttt{E615}}            &15 \%                 &$1.160\pm0.020$&$206/140$\\
		\href{https://doi.org/10.17182/hepdata.15988.v2}{\texttt{NA10} $(194\GeV)$}&6.4\%                 &$0.997\pm0.014$&$107/ 67$\\
		\href{https://doi.org/10.17182/hepdata.30229   }{\texttt{NA10} $(286\GeV)$}&6.4\%                 &$0.927\pm0.013$&$ 95/ 73$\\
		\href{https://doi.org/10.17182/hepdata.15649.v2}{\texttt{WA70}}            &\textcolor{blue}{32\%}&$0.737\pm0.012$&$ 64/ 99$\\
		\hline\hline
	\end{tabular}
\end{table}
For the uncertainties from nuclear PDFs and $\alS$, the coefficients $\ga_{i\al}$ are estimated as derivatives of the theory predictions with respect to $\alS$ and the uncertainty eigenvectors of the nuclear PDFs as provided by the \texttt{nCTEQ15} set.
This linear approximation is valid only when the minimisation parameters are close to their optimal values.
It was verified that this condition was satisfied for the performed fits.
\par
The uncertainty of the perturbative calculation is estimated by varying the renormalization scale $\mu_R$ and factorization scale $\mu_F$ by a factor of two up and down, separately for $\mu_R$ and $\mu_F$.
The scales were varied using \texttt{APPLgrid}, and the variations were coherent for all data bins.
Renormalization scale variation for DGLAP evolution was not performed.
We observe a significant dependence of the predicted cross-sections on $\mu_R$ and $\mu_F$: the change in predictions is $\sim10\%$, which is comparable to the normalization uncertainty of the data.
This dependence indicates that next-to-next-to-leading order corrections may be significant.
\par
In order to estimate the uncertainty related to the flexibility of chosen parameterisation, the fit is repeated with a varied initial scale $Q_0^2=1.9\pm0.4\GeV^2$.
This variation leads to only a small change in $\chi^2$ ($\Delta\chi^2\lesssim1$).
In order to stay below the charm mass, for variation up to $Q^2_0=2.3\GeV^2$ the mass threshold $m_c^2$ was shifted up by the same amount.
The effect of such a change in the charm mass threshold by itself was found to be negligible.

\section{Results}\label{sec:results}
\begin{figure*}[tb]
	\centerline{\includegraphics[width=18cm]{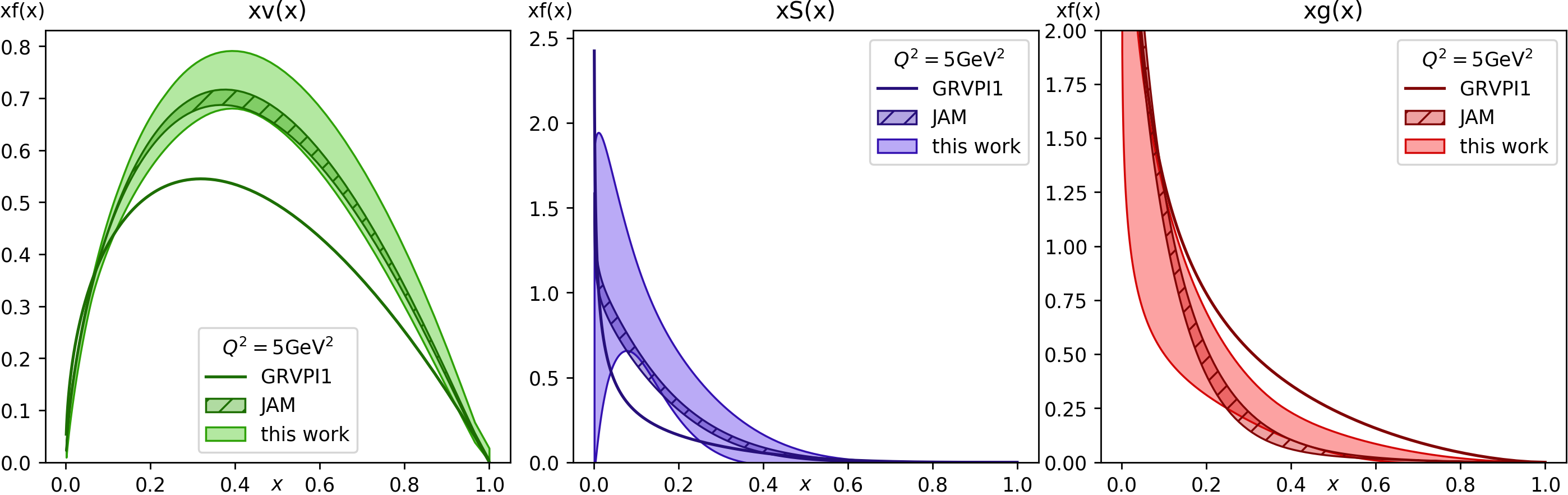}}
	\caption{\label{fig:comparePDFs}
	Comparison between the pion PDFs obtained in this work, a recent determination by
	the JAM collaboration\cite{JAM},
	and the \texttt{GRVPI1} pion PDF set\cite{GRVPI}.
	}
\end{figure*}
\begin{figure}[tb]
	\centering
	\includegraphics[width=0.4\textwidth]{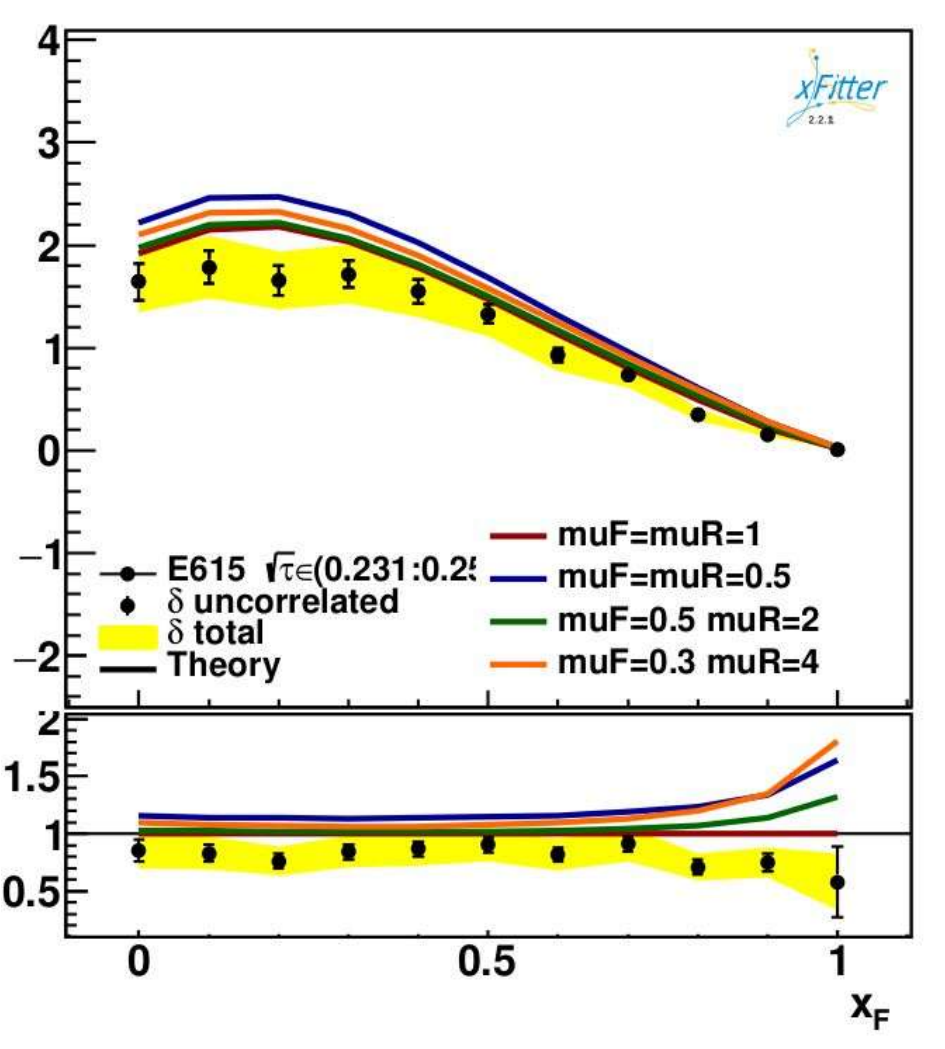}
	\caption{\label{fig:scaleVar}
We display the scale variation of the cross section for a sample E615 $\sqrt{\tau}$ bin as a function of $x_F$.
Note, the normalization factor of Table~\ref{tbl:data} ($1.60\pm 0.020$) has not been applied.
We observe the relative impact of the scale variation is minimal except
at very large $x$ ($x\gtrsim 0.9$).
	}
\end{figure}


\Figref{comparePDFs} shows the obtained pion PDFs in comparison to
a recent analysis by JAM\cite{JAM}, and to \texttt{GRVPI1}\cite{GRVPI}.
The new valence distribution presented here is in good agreement with JAM,
and both differ with the early GRV analysis.
The relatively difficult to determine sea and gluon distributions are different in all three PDF sets,
however, this new PDF and the JAM determination agree within the larger uncertainties of our fit.

In the case of valence distribution, the dominant contribution to the
uncertainty estimate is the variation of the scales $\mu_R$ and
$\mu_F$.
For the sea and gluon distributions, the missing fragmentation
contribution to prompt photon production is the dominant uncertainty
source, and the effect of scale variation is also significant.
Recall that JAM used the E615 and NA10 DY data (as we did),
but used the HERA leading neutron electroproduction data while we used a direct photon analysis 
with a large normalization uncertainty ({\it c.f.}, Table~\ref{tbl:data}).

\begin{figure*}[p]
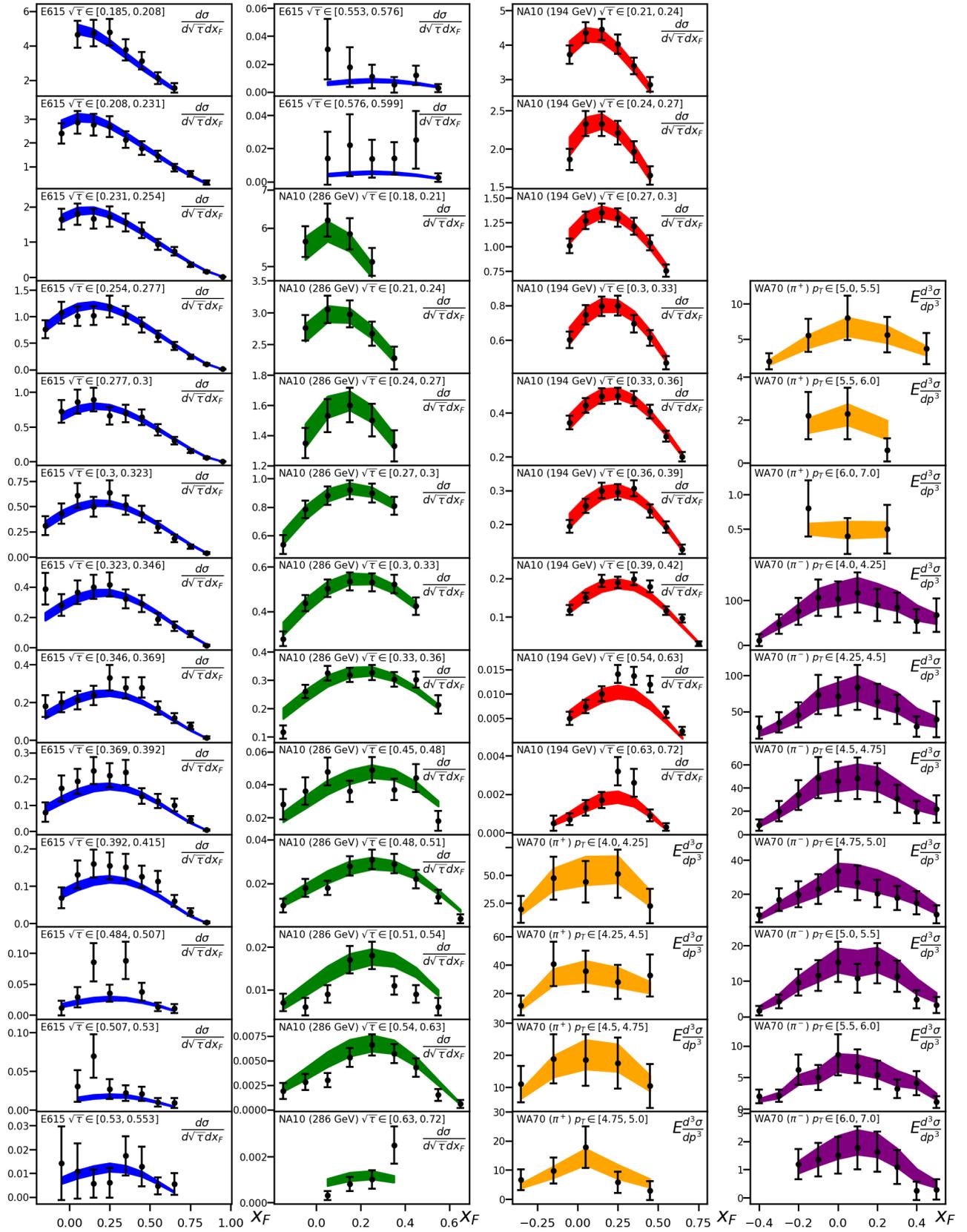

	\adjustimage{max width=\linewidth,max height=\paperheight,center}{FittedData.png}
	\caption{\label{fig:FittedData}
	Considered experimental data and corresponding theory predictions.
	The displayed theory predictions include correlated shifts.
	Bands of different colors correspond to different datasets.
	Width of the bands shows uncertainty of the theory predictions.
	The cross-sections are shown in the same format as adopted by corresponding experimental papers.
	The \texttt{E615} data is given as
	$\dif^2\sigma/(\dif\sqrt\tau\dif x_F)$ in nb/nucleon, averaged over each $(\sqrt\tau, x_F)$ bin.
	The DY data from the \texttt{NA10} experiment is
	$\dif^2\sigma/(\dif\sqrt\tau\dif x_F)$ in nb/nucleus, integrated over each $(\sqrt\tau, x_F)$ bin.
	The \texttt{WA70} data on direct photon production is given as invariant cross-section
	$E\dif^3\sigma/\dif p^3$ in pb, averaged over each $(p_T, x_F)$ bin.
	}
\end{figure*}

A comparison between experimental data and theory predictions obtained
with the fitted PDFs is presented in \figref{FittedData}.  Reasonable
agreement between data and theory is observed, with no systematic
trends for any of the kinematic regions.

\begin{table}[!t]
	\caption{\label{tbl:momentum}
	Momentum fractions of the pion carried by the valence, sea and gluon PDFs at different scales $Q^2$
	as determined in this work in comparison to other studies.
	}
	\centering
          \setlength{\tabcolsep}{5pt}
	\begin{tabular}{lllll}
		\hline\hline
		&$\langle xv\rangle$&$\langle xS\rangle$&$\langle xg\rangle$&\hspace{-1mm}\begin{tabular}{@{}c@{}}$Q^2$\\$(\text{GeV}^2)$\end{tabular}\hspace{-1mm}\\
		\hline\hline
		JAM\cite{JAM}              &$0.54 \pm0.01 $&$0.16\pm0.02$&$0.30\pm0.02$&1.69\\
		JAM (DY)                   &$0.60 \pm0.01 $&$0.30\pm0.05$&$0.10\pm0.05$&1.69\\
		this work                  &$0.55 \pm0.06 $&$0.26\pm0.15$&$0.19\pm0.16$&1.69\\
		\hline
		Lattice-3\cite{Lattice2016}&$0.428\pm0.030$&&                          &4\\
		SMRS\cite{SMRS}            &$0.47         $&&                          &4\\
		Han et al.\cite{FitHan}    &$0.51 \pm0.03 $&&                          &4\\
                %
		GRVPI1\cite{GRVPI}         &$0.39         $&$0.11$       &$0.51$       &4\\
		Ding et al.\cite{Ding:2019lwe}&$0.48 \pm0.03 $&$0.11\pm0.02$&$0.41\pm0.02$&4\\
		this work                  &$0.50 \pm0.05 $&$0.25\pm0.13$&$0.25\pm0.13$&4\\
		\hline
		JAM                        &$0.48 \pm0.01 $&$0.17\pm0.01$&$0.35\pm0.02$&5\\
		this work                  &$0.49 \pm0.05 $&$0.25\pm0.12$&$0.26\pm0.13$&5\\
		\hline
		Lattice-1\cite{Lattice1997}&$0.558\pm0.166$&&                          &5.76\\
		Lattice-2\cite{Lattice2003}&$0.48 \pm0.04 $&&                          &5.76\\
		this work                  &$0.48 \pm0.05 $&$0.25\pm0.12$&$0.27\pm0.13$&5.76\\
		\hline
		WRH\cite{Wijesooriya}      &$0.434\pm0.022$&&                          &27\\
		ChQM-1\cite{ChQM_2012}     &$0.428        $&&                          &27\\
		ChQM-2\cite{ChQM_2018}     &$0.46         $&&                          &27\\
		this work                  &$0.42 \pm0.04 $&$0.25\pm0.10$&$0.32\pm0.10$&27\\
		\hline
		SMRS\cite{SMRS}            &$0.49 \pm0.02 $&&                          &49\\
		this work                  &$0.41 \pm0.04 $&$0.25\pm0.09$&$0.34\pm0.09$&49\\
		\hline\hline
	\end{tabular}
\end{table}

\begin{figure*}[tb]
	\centerline{\includegraphics[width=18cm]{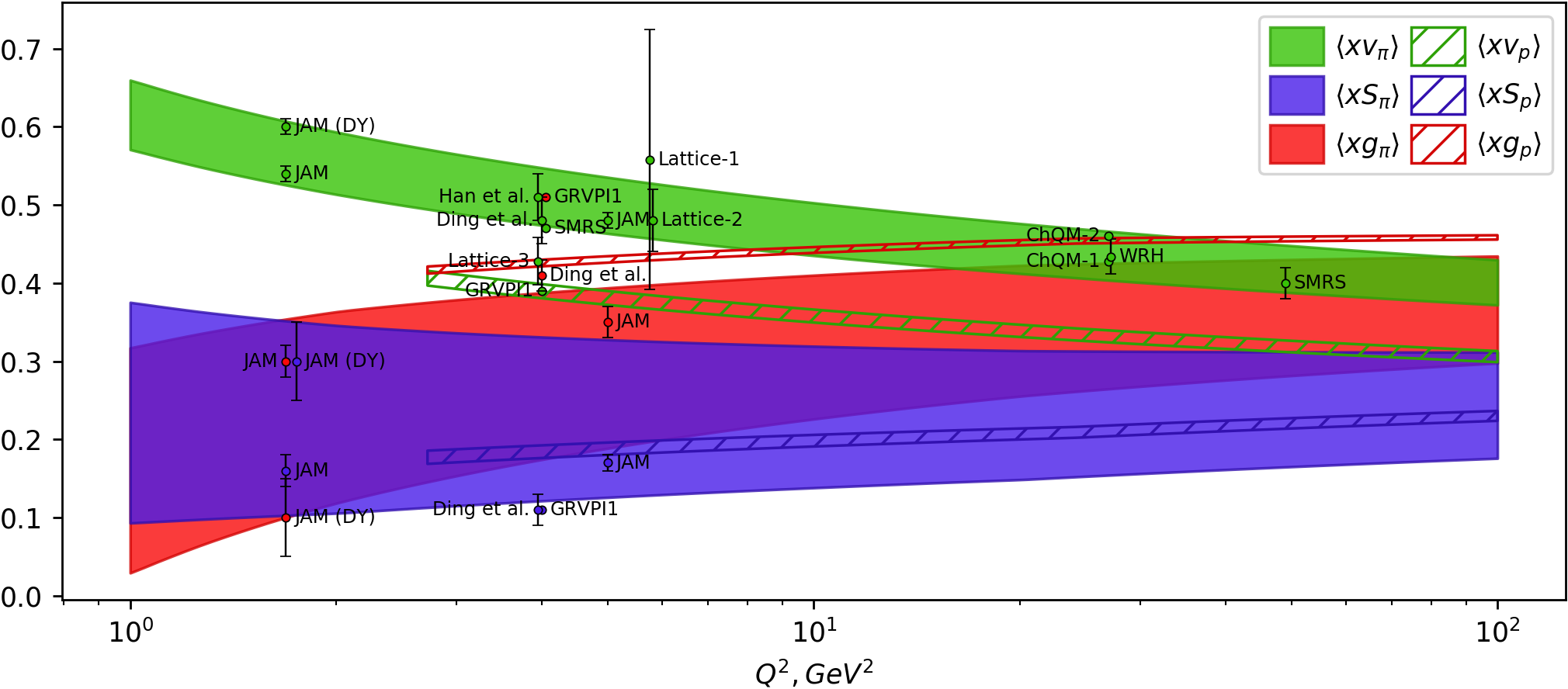}}
	\caption{\label{fig:evolvingFractions}
Momentum fractions of the pion as a function of $Q^2$.
The error bands include all uncertainties described in Section \ref{sec:errors}.
Analogous momentum fractions in the proton PDF set \texttt{NNPDF31\_nlo\_as\_0118} are shown for comparison.
The labeled green, red, and blue points show respectively valence, sea, and gluon momentum fractions as reported by other studies.
The references and numerical values for these points are listed in \tblref{momentum}.
	}
\end{figure*}



We now examine the high-$x$ behavior of the valence PDF.
The asymptotic limit of the valence PDF as $x\to 1$ has been studied
extensively in the literature. 
For example, 
Nambu-Jona-Lasinio models\cite{HoltRoberts} favor a  $v(x)\sim(1-x)$ behavior
while  approaches based on the Dyson-Schwinger equations (DSE)\cite{DSE_2018,DSE_2019}
obtain a very different  $v(x)\sim(1-x)^2$.
The discrepancy between DSE predictions and fits to pion Drell-Yan
data is well-known\cite{Wijesooriya, SoftGluonResummation,DSE_2019},
and  it has been demonstrated that soft-gluon threshold resummation
(which was not included in this analysis) may be used to account for this
disagreement\cite{SoftGluonResummation}.
Alternatively, DSE calculations using inhomogeneous Bethe-Salpeter
equations\cite{DSE_2019} 
can produce PDFs consistent with the linear
behavior of the $v(x)$ in the region covered by DY data, pushing the
onset of the $(1-x)^2$ regime to very high $x$.

Although the asymptotic behavior of the valence PDF is a theoretically interesting measurement,
we will explain in the following why we are unable to determine this with the current analysis;
conversely, details of the asymptotic region therefore do not impact our extracted pion PDFs.


First, the asymptotic DSE results only apply at asymptotically large $x$ values.
While the precise boundary is a subject of debate, Ref.~\cite{DSE_2019} demonstrates
that the perturbative QCD predictions may only set in very near $x=1$; hence, the observed
$(1-x)^1$ behavior could be real where the data exists.
Consequently, it is entirely possible to have $(1-x)^1$ behavior at intermediate to large $x$,
but then still find $(1-x)^2$ asymptotically.
Except for the threshold-resummed calculation of Ref.~\cite{SoftGluonResummation},
the fits to the E615 and NA10 data\cite{E615,NA10,NA10revised,GRS1997,GRS1999,JAM}
generally obtain high-$x$ behaviors that are closer to $(1-x)^1$ than the DSE result. 
This explains how these many fits can coexist with the asymptotic DSE limit. 


What would it take to be able to accurately explore the $x\to1$ asymptotic region?
This region is challenging both experimentally and theoretically.
On the experimental side, in the limit $x\to1$ the PDFs are rapidly decreasing. 
Hence the cross section is very small,  making large $x$ measurements
difficult. Fig.~\ref{fig:FittedData} displays the full set of data we fit, and it is evident
that the number of data at the largest $x_F$ values is limited. 
The issues on the theoretical side are also complex.
In Fig.~\ref{fig:scaleVar} we present the scale
dependence for a sample subset of the E615 data. We see the relative
scale dependence across the $x_F$  kinematic range is generally under
control, with the exception of the very large $x_F$ limit;
hence, the theoretical uncertainties of the NLO calculation increase
precisely in the region required to extract the asymptotic
behavior.
Therefore,  we reiterate
that this analysis does not possess sufficient precision
to infer definitive conclusions on the asymptotic  $x\to 1$ limit of
the pion structure function.


Furthermore, to properly study the $x\to1$ asymptotic limit,
a more sophisticated parametric form is required.
The polynomial form for the pion valence PDF of Eq.~\ref{eq:parm}
has only two or three free parameters  $\{B_v, C_v, D_v\}$, and the large $x$ behavior is
dominantly controlled by the $C_v$ coefficient.
In light of the results of Ref.~\cite{DSE_2019}, a more flexible
parameterization is required to accommodate separate $x$-dependence
at both intermediate to large $x$ and then the asymptotic region.


The threshold resummation calculation\cite{Laenen_2001} has generated significant interest, in part, because
the resulting pion PDFs had a valence structure closer to the DSE \mbox{$(1-x)^2$} form.\cite{SoftGluonResummation}
However it is important to recall that the PDFs themselves are not physical observables,
but depend fundamentally on the underlying schemes and scales used for the calculation.
If scheme-dependent PDFs are used with properly matched scheme-dependent hard cross sections,
the result will yield scheme-independent observables as in Fig.~\ref{fig:FittedData}.
Additionally, were we to perform our analysis with the threshold resummed scheme,
it would be most appropriate to do this for all the processes including both
the DY and direct photon processes; however PDFs obtained
with resummation corrections would also require resummed hard cross sections
for the predictions.
In contrast, our NLO analysis effectively absorbs resummation
corrections (approximately) into the PDFs; but, it can be used to predict cross
sections at NLO for future experiments using existing NLO open source tools.


To study the restrictions of our parameterization, 
we introduce an addition parameter $D_v$ for  our valence PDF.
This term has an impact on the intermediate to large $x$ behavior
as evidenced by the change on the $C_v$ parameter, {\it c.f}, Table~\ref{tbl:parameters}.
However the improvement in the $\chi^2/N_{DoF}$ is minimal (1.19 {\it vs} 1.18), 
as is the $x$-dependence as shown in Fig.~\ref{fig:HighX}.


\figref{evolvingFractions} shows the obtained momentum fractions in the pion as a function of $Q^2$.
Recall that  $A_S$ is a fit variable and $\{A_v,A_g\}$ are determined by the sum rules of \eqnref{sumrules}.
Above the charm and bottom mass thresholds ($Q>m_c, m_b$),
the $c$ and $b$ quarks and anti-quarks are included in the sea distribution.
For comparison, we have overlaid the results from the other studies listed  in \tblref{momentum};
these results are consistent within our uncertainties, except for the lattice simulation of
Ref.~\cite{Lattice2016} (denoted by label "Lattice-3") and the GRVPI1 set.


Additionally, we have displayed the proton momentum fractions for the 
\texttt{NNPDF31\_nlo\_as\_0118}\cite{NNPDF} set.
Relative to the proton,  we find the valence of the pion is larger, 
the gluon  is smaller,  and the sea component is similar, within uncertainties. 
We also note the \hbox{$Q^2$-dependence} of the various components are similar for both the pion and the proton,
as they are all determined by the same DGLAP evolution equations.

\section{Summary and outlook}

We have presented the first open-source analysis of pion PDFs.
We have used Drell-Yan and prompt photon production data 
with  \texttt{APPLgrids} generated from \texttt{MCFM} to extract the PDFs at NLO.
Additionally, we have performed a complete analysis of both experimental and theoretical uncertainties
including renormalization and factorization scale variation ($\mu_R, \mu_F$),
strong coupling variation ($\alpha_S$),
and PDFs (both proton and nuclear).

Comparing with other pion PDFs from the literature,
our results are similar to  JAM, but differ from the GRVPI1 set.
Although the valence distribution is comparably well constrained, the
considered data are not sensitive enough to unambiguously determine
the sea and gluon distributions.
We note our uncertainties are larger than JAM due to
i)~the theoretical uncertainties discussed above, and
ii)~the large normalization uncertainty on our direct photon analysis
(JAM uses  LN electroproduction instead).
This is an area where new data, such as $J/\Psi$ production, 
could play an important role in constraining the gluon.\cite{Chang:2020rdy}

The data are reasonably
well-described by NLO QCD, but the sensitivity to $\mu_R$ and $\mu_F$
indicates that next-to-next-to-leading order corrections could be
significant, especially in the very large $x$ region;
this precludes us from extracting the asymptotic behavior
of the valence distribution.

We will provide the  extracted pion PDFs in the \texttt{LHAPDF6} PDFs library, 
and the \texttt{APPLgrid} grid files in the \texttt{Ploughshare}\cite{Ploughshare} grid library.
%
%
Since  xFitter is an open-source program, it provides
the community with a versatile tool to study meson PDFs
which can be extended to perform new analyses.
In particular, when data from future experiments, such
as COMPASS++/AMBER,[50] becomes available, studying
more flexible parameterization forms and including
corrections beyond NLO will be of interest.


\begin{acknowledgments}
The authors would like to thank the DESY IT department for the provided computing resources and for their support of the \xFitter developers.
We thank
Tim~Hobbs,
Pavel~Nadolsky,
Nobuo~Sato,
Ingo~Schienbein,
and
Werner~Vogelsang
for useful discussions.
A.K.\  acknowledges the support of Polish National Agency for Academic Exchange (NAWA) within The Bekker Programme,
and F.O.\ acknowledges US DOE grant  DE-SC0010129.
\end{acknowledgments}
\cleardoublepage
\bibliography{references}
\end{document}